\documentclass[conference]{IEEEtran}

\usepackage{cite}
\usepackage{amsmath,amssymb,amsfonts}
\usepackage{algorithmic}
\usepackage{graphicx}
\usepackage{textcomp}
\usepackage{xcolor}
\def\BibTeX{{\rm B\kern-.05em{\sc i\kern-.025em b}\kern-.08em
    T\kern-.1667em\lower.7ex\hbox{E}\kern-.125emX}}

\usepackage[utf8]{inputenc}
\usepackage{url}
\usepackage{tikz}
\usepackage{eso-pic}
\usepackage[colorlinks,allcolors=black,urlcolor=blue]{hyperref}

\AddToShipoutPicture{%
  \sffamily\scriptsize
  \setlength\unitlength{1mm}%
  \AtPageUpperLeft{%
    \put(40,-3){\parbox[t]{0.75\textwidth}{%
        \centering
        This article has been accepted for publication in
        the International Conference on Transparent Optical Networks (2025)
        as an Invited paper.  This is the author's version.
        Citation information:
        \href{https://dx.doi.org/10.1109/ICTON67126.2025.11125476}{%
          DOI 10.1109/ICTON67126.2025.11125476}
    }}%
  }%
  \AtPageLowerLeft{%
    \put(40,3){\parbox[b]{0.8\textwidth}{%
        \tiny
        \centering
        © 2025 IEEE. Personal use is permitted, but
        republication/redistribution requires IEEE permission.
        See \url{https://www.ieee.org/publications/rights/index.html}
        for more information.
    }}%
  }%
}

\begin{document}

\title{Quantum-Aware Network Planning and Integration}

\author{\IEEEauthorblockN{Cédric Ware}
\IEEEauthorblockA{\textit{LTCI, Télécom Paris,} \\
\textit{Institut Polytechnique de Paris,}\\
Palaiseau, France \\
0000-0001-5957-5632}
\and
\IEEEauthorblockN{Mounia Lourdiane}
\IEEEauthorblockA{\textit{SAMOVAR, Télécom SudParis,} \\
\textit{Institut Polytechnique de Paris,}\\
Évry, France \\
0000-0003-2459-685X}
}

\makeatletter \def\@IEEEpubid{\parbox[t]{0.9\textwidth}{%
    © 2025 IEEE.  Personal use of this material is permitted.
    Permission from IEEE must be obtained for all other uses, in any
    current or future media, including reprinting/republishing this
    material for advertising or promotional purposes, creating new
    collective works, for resale or redistribution to servers or
    lists, or reuse of any copyrighted component of this work in other
    works.
}} \@IEEEusingpubidtrue
\def\IEEEpubidadjcol{\if@IEEEusingpubid\enlargethispage{-\@IEEEpubidpullup}\fi}
\makeatother

\maketitle

\begin{abstract}
  In order to broaden the adoption of highly-demanded quantum
  functionalities such as QKD, there is a need for having quantum
  signals coexist with classical traffic over the same physical medium,
  typically optical fibers in already-deployed networks.  Beyond the
  experimental point-to-point demonstrations of the past few years,
  efforts are now underway to integrate QKD at the network level:
  developing interfaces with the software-defined-network ecosystem;
  but also network planning tools that satisfy physical-layer contraints
  jointly on the classical and quantum signals.  We have found that in
  certain situations, naïve network planning prioritizing quantum traffic
  drastically degrades classical capacity, whereas a quantum-aware
  wavelength assignment heuristic allows coexistence with minimal impact
  on both capacities.  More such techniques will be required to enable
  widespread deployment of QKD and other future quantum functionalities.
\end{abstract}


\section{Introduction}

The ongoing development of quantum technologies, after first revolutionizing
science then society over the past decades thanks to electronic and
opto-electronic components, may be on the cusp of a
\emph{second quantum revolution}
\cite{dowling-JSTOR-2003-second-q-revolution}
in at least several fields such as
computing, communications, metrology and, strategically, security.
Quantum key distribution (QKD) is one of the most prominent
applications, offering physics-based secure communications that are
resistant to the threats posed by future quantum computers. To bring
these technologies to a wide range of end-users, the realization of
a large-scale quantum network, possibly culminating in a full
quantum Internet, will be necessary.
These networks are envisioned to evolve in distinct stages
\cite{wehner-Science-2018-quantum-internet}.  The first stage involves the
exchange of quantum states through trusted relays, allowing QKD
between distant nodes if intermediate relays are trusted;
several city-scale
\cite{peev-NJP-2009-secoqc-vienna,sasaki-OpEx-2011-tokyo-qkd,chen-QI-2021-q-man,martin-QI-2024-madqci,bersin-PRA-2024-boston-fiber-q-network}
and even country-scale
\cite{mao-OE-2018-qkd-backbone,chen-JPhys-2021}
experimental networks have already demonstrated this stage.
Further stages involve the distribution of entangled quantum states
and their storage in quantum memories, enabling quantum teleportation
and more advanced protocols as well as metrology applications improving
performance over classical systems \cite{gottesman-PRL-2012-long-baseline-q-repeaters,degen-RMP-2017-q-sensing};
ultimately, full error-corrected quantum-computing networks
will allow protocols requiring multiple round-trips or even
more complex exchanges, enabling applications
such as secure multi-party quantum computing.

However, whatever their stage, the commercial viability of these networks
will require sharing the communication infrastructure with existing
classical networks, rather than relying on dedicated (``dark'')
optical fibers and ending up having to deploy a whole new infrastructure.
Encouragingly, quantum communications have been demonstrated over
the same fiber as classical traffic
\cite{kumar-2015-coexistence,thomas-Optica-2024-q-teleportation-coexistence},
including symbiotically using
the classical traffic as a phase reference for the quantum transmission
\cite{aymeric-OFC-2022-symbiotic}.
Yet there may be a limit to the amount of classical traffic that
can be tolerated by this coexistence, in which case
we have found that prioritizing quantum traffic can sometimes
``clog'' network links to classical traffic
\cite{ware-PTL-2022-impact},
negating any commercial viability.
Thus, quantum-classical coexistence at scale will require not only technological
improvements in quantum repeaters and other devices \cite{azuma-RMS-2023}
but also network-wide strategies and tools for control
\cite{sebastian-lombrana-ArXiv-2024-blueprint-q-network,iqbal-JOCN-2025-sdn-cv-qkd-coexistence}
and planning \cite{garbhapu-ecoc-2024,pouryousef-QuNet-2023-q-network-planning}.

In the following sections, this paper will focus on the physical
issues and then the network planning issues limiting quantum
communications in a network context, then give an overview of our
ongoing simulation studies of the impact on classical network capacity
of coexisting with CV-QKD.

\IEEEpubidadjcol
\section{Physical issues limiting quantum communications in a network context}
\label{sec/phy-issues}

The core function of a quantum network is the transmission of
delicate quantum states (possibly carried by single photons).
An arbitrary state cannot be copied due to the no-cloning
theorem, meaning that the signals carrying them cannot be
amplified \cite{wei-LPR-2022-real-world-quantum-networks}.
Quantum signals are thus inherently weak and highly susceptible
to noise, especially when coexisting with powerful classical signals
that generate crosstalk through linear and nonlinear processes.
Performance metrics such as the quantum bit error ratio (QBER)
or the secure key rate (SKR) for QKD are significantly affected
by this coexistence.

\subsection{Attenuation limits range}
\label{sec/phy/attenuation}

Attenuation in standard singlemode fiber (SSMF), typically
around 0.2~dB/km, severely limits the range of
quantum communications: the SKR is strongly dependent on
attenuation, leading to a drop-off in performance
with increasing distance \cite{laudenbach-cvqkd-theory-2018}.
Discrete-variable (DV) QKD systems have demonstrated operation over
several hundred kilometers (up to 1002~km in \cite{liu-QF-2023-q-1002km},
albeit only at $\sim 10^{-3}$~key~bits per seconds),
but continuous-variable (CV) QKD,
that could be cheaper because it can be realized with coherent-detection
equipment similar to that of classical systems, is
generally constrained to tens of kilometers (recently reaching 100~km
\cite{hajomer-Science-2024-cv-q-llo-100km}).
This limitation precludes
the use of transparent routing common in classical optical networks,
necessitating opaque routing with repeaters, which in turn requires
trusting intermediate nodes—at least in the first stage of the
quantum Internet;
in future stages, being able to distribute entangled quantum states
will allow quantum protocols that are secure even with untrusted
intermediate nodes \cite{wehner-Science-2018-quantum-internet}.

Hollow-core fibers (HCF) offer a potential solution: they seek
to lower attenuation by having the signal propagate mostly in air
rather than silica, and have recently caught up with SSMF
\cite{jasion-OFC-2022-hcf-dnanf}, with reported losses as low
as 0.1~dB/km \cite{petrovich-ArXiV-2025-hcf}.
However, HCF technology is less mature and not yet deployed,
thus not solving the problem of integrating QKD into existing networks.

\subsection{Decoherence limits latency}
\label{sec/phy/decoherence}

Quantum systems are also vulnerable to decoherence
\cite{wehner-Science-2018-quantum-internet},
a process by which
superpositions of quantum states become entangled with the environment,
effectively collapsing into statistical mixtures of quasi-classical states.
This affects network latency requirements, in that longer propagation times
(for quantum communications) or storage durations (for memories) increase
the likelihood of decoherence.
Quantum protocols involving multi-step exchanges or requiring
synchronization between distant parties may exacerbate this issue,
as the added latency will drive up storage time requirements.

\subsection{Crosstalk (linear and nonlinear) limits coexistence}
\label{sec/phy/crosstalk}

Beyond the intrinsic limitations described above, the coexistence
of classical and quantum traffic in the same optical-fiber links
introduces additional challenges, primarily related to imperfect
isolation and optical nonlinearity.

As for any kind of multiplexing, quantum and classical signals
must be carried over separate propagation modes—the word ``modes''
being used here in the general sense, such as wavelength (or even
wavebands, e.g.\ using the C band for classical and the O band
for quantum channels), polarization, or time slots; or spatial modes
(multicore \cite{lin-ACCESS-2020-compatibility-qkd}, possibly multimode)
over fibers other than SSMFs.
Linear crosstalk occurs when the demultiplexing device or system
imperfectly separates these modes.
For wavelength-division multiplexing (WDM), at constant device
quality, increasing channel spacing improves isolation.
Cumulating multiple techniques also improves isolation,
as multiple filtering devices will be cascaded, e.g.
a WDM demultiplexer in series with a polarization beam splitter.

In addition, optical nonlinearity couples modes together,
notably four-wave mixing (FWM) driven by the Kerr effect,
and spontaneous Raman scattering (SpRS).
Photons from powerful signals (at least two signals for FWM,
or one signal coupled with vibration modes in the fiber's material
for SpRS) generate frequency-shifted photons that can end up
in the same mode as quantum signals, precluding separation
by filtering.  Both these effects are exacerbated by high
optical power.  FWM can be mitigated, as with linear crosstalk,
by increasing WDM channel spacing up to a certain point
(e.g.\ 200~GHz in \cite{zavitsanos-ICTON-2019-dv-qkd-c-band});
less so for SpRS which has a much broader spectral response
($\sim 15$~THz in silica fiber \cite{hollenbeck-JOSAB-2002-raman-silica,bromage-JLT-2004-raman}),
thus ending up being the dominant
nonlinear noise source.
(Here also, HCFs could be an interesting solution,
as SpRS generation can be several tens of dB lower
than in SSMF \cite{honz-JLT-2023-c+l-band-dv-qkd}.)

From a network operator point of view, not only these issues
would degrade quantum traffic performance in the presence of
classical signals; they could conversely impose restrictions
on the classical traffic, such as limiting the optical power
levels or the number of WDM channels, so as to be able to
carry the quantum traffic at the required performance levels.
This will be mentioned again in section~\ref{sec/net/joint-planning}.

\section{Network planning issues for deploying quantum communications}
\label{sec/net-issues}

\subsection{Management of quantum communication devices}
\label{sec/net/management}

From a network operator standpoint, a network functionality must be
not only present in hardware but also integrated into the network
management system (NMS) software.  The NMS must be aware of
nodes and links with quantum capabilities, allocate quantum channels,
configure and monitor devices (e.g.\ via software-defined networking
(SDN) protocols).
It may also secure its own control plane by QKD, as well as provide
QKD-generated keys as a service to end-users through a key management
system.
Sustained efforts to elaborate these aspects, including
standardization so as to enable multi-vendor interoperability, have
been supported by several projects such as Europe’s current Quantum
Secure Networks Partnership (QSNP) \cite{QSNP-website}. Ongoing
development of a detailed framework for control and orchestration
is described in
\cite{sebastian-lombrana-ArXiv-2024-blueprint-q-network}.

\subsection{Joint optimization of classical and quantum traffic}
\label{sec/net/joint-planning}

As mentioned in section~\ref{sec/phy/crosstalk}, the impact between
quantum and classical coexistence can be considered both ways:
degrading quantum performance directly, or limiting classical network
capacity indirectly when attempting to provide performance guarantees
on the quantum channels.
Network planning must therefore balance these competing demands to
jointly optimize resource allocation for classical and quantum traffic.
This may require an accurate model of the physical propagation,
especially crosstalk and nonlinear noise generation;
such a model may be used directly, or embedded into physical
parameters that can then be exposed to an SDN-based network
architecture, as suggested in
\cite{iqbal-JOCN-2025-sdn-cv-qkd-coexistence}.

\section{Results and discussion of our ongoing simulation studies}
\label{sec/results}

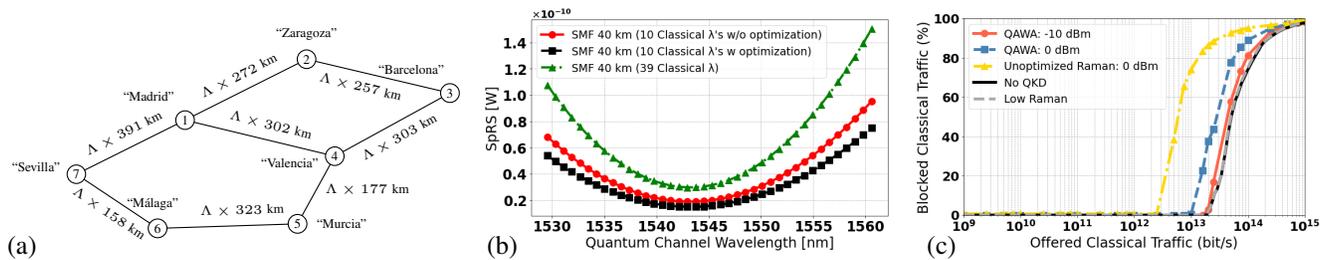
\begin{figure*}
  \centering
  \begin{tikzpicture}[scale=0.6,font=\tiny]
    \coordinate (Madrid) at (-3.6919444,41.3);
    \coordinate (Barcelona) at (2.1769444,41.9);
    \coordinate (Valencia) at (-0.375,40.5);
    \coordinate (Sevilla) at (-6.1,40.1);
    \coordinate (Zaragoza) at (-1,42.65);
    \coordinate (Malaga) at (-4.3,38.9);
    \coordinate (Murcia) at (-1.2,39.0);
    \draw (Barcelona)
    -- node [below,sloped] {$\Lambda\times 303$~km} (Valencia)
    -- node [right] {$\Lambda\times 177$~km} (Murcia)
    -- node [above,sloped,xshift=2mm] {$\Lambda\times 323$~km} (Malaga)
    -- node [below,sloped] {$\Lambda\times 158$~km} (Sevilla)
    -- node [above,sloped] {$\Lambda\times 391$~km} (Madrid)
    -- node [above,sloped] {$\Lambda\times 272$~km} (Zaragoza)
    -- node [below,sloped,xshift=-2mm] {$\Lambda\times 257$~km} (Barcelona)
    (Madrid) -- node [above,sloped,xshift=1mm] {$\Lambda\times 302$~km} (Valencia);
    \foreach \id/\o/\r/\c/\t in {%
      1/above left/100/Madrid,3/above left/70/Barcelona,%
      4/left/225/Valencia,7/left/135/Sevilla,%
      2/above/north/Zaragoza,%
      6/above/120/Malaga/Málaga,5/right/east/Murcia} {
      \node [draw,circle,fill=white,inner sep=1pt] (node \c) at (\c) {\id};
      \node [\o] at (node \c.\r) {``\t''};
    }
  \end{tikzpicture}\hfil
  \includegraphics[width=0.3\linewidth]{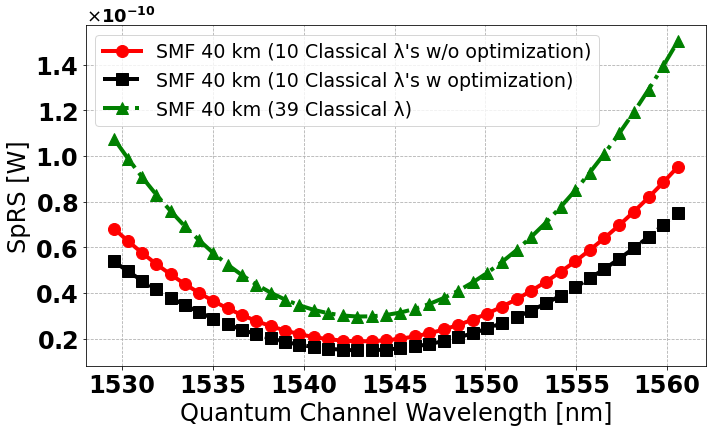}\hfil
  \includegraphics[width=0.3\linewidth]{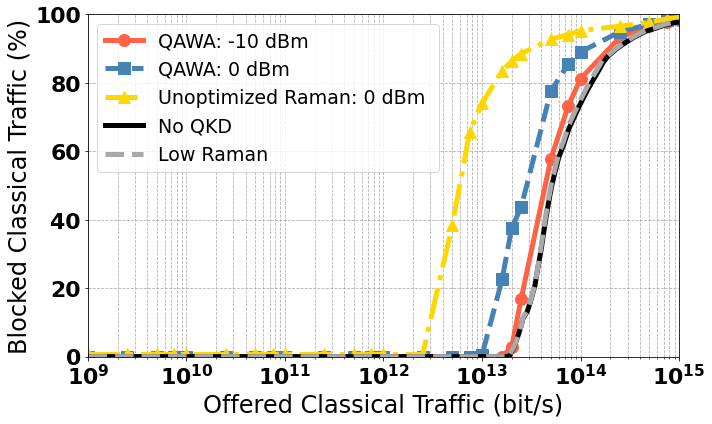}
  \\
  \vspace{-\baselineskip}
  \quad(a)\hspace{0.33\linewidth}(b)\hspace{0.3\linewidth}(c)\hfill\hfill
  \caption{%
    (a) Network topology with 7 nodes and 8 bidirectional links,
    adapted from \cite{pavon-2015-net2plan},
    representing a fictitious mesh network connecting the 7
    most populated cities in Spain, with distances uniformly
    scaled by a factor $\Lambda$ (ranging from $10^{-2}$ to $10^{-1}$).
    (b) Impact of the quantum channel placement within the
    C~band on the amount of SpRS noise generated by surrounding
    classical channels.
    (c) Proportion of classical traffic blocked as a function
    of traffic offered, with a CV-QKD traffic of 280~Mbit/s network-wide
    with $12$~\% margins, with or without the QAWA heuristic and with
    $0$ or $-10$~dBm per classical channel;
    the red curve (all optimizations) is virtually identical
    to the reference black curve (no coexistence, classical traffic only).
    Previously published in \cite{garbhapu-ecoc-2024}.
    \label{fig/our-study}
  }
\end{figure*}

\subsection{Problem statement}
\label{sec/results/overview}

We conducted simulations on a small-topology network with links
sharing CV-QKD and classical traffic via WDM, all in the C band.
The topology, shown in Fig.~\ref{fig/our-study}(a),
was based on a fictitious example network from Net2Plan
\cite{pavon-2015-net2plan}, with distances uniformly scaled down to
reflect feasible CV-QKD ranges.
We modeled SKR as a function of attenuation (due to propagation length)
and SpRS (from the classical channels) according to
\cite{laudenbach-cvqkd-theory-2018} for SKR and
\cite{bromage-JLT-2004-raman} for SpRS.
All nodes were treated as trusted relays aggregating QKD traffic,
and transparent for classical traffic.
Since QKD is more fragile, we prioritized its allocation; and afterward
allocated transparent lightpaths for classical traffic, ensuring that
optical noise did not disrupt QKD traffic.
We then evaluated the blocking probability, that is, the ratio of
successfully-allocated traffic over the total amount of offered traffic.

\subsection{Findings}
\label{sec/results/findings}

\subsubsection{Placement of classical and CV-QKD channels within the C band:
  quantum-aware wavelength assignment (QAWA)}
\label{sec/results/channel-placement}

The allocation of the CV-QKD channel within the C band can be
optimized to minimize SpRS from other channels. Our simulations
suggest placing the CV-QKD channel near the middle of the C band
can reduce SpRS, as shown in Fig.~\ref{fig/our-study}(b)
\cite{garbhapu-ecoc-2024}. However, another study
using a different SpRS model suggests placing the CV-QKD channel
on the high-frequency side of the C band
\cite{iqbal-JOCN-2025-sdn-cv-qkd-coexistence},
indicating that performance is strongly dependent on the SpRS
model used and that further investigation is needed.

Assuming an accurate SpRS model, we can take a further step of
dynamically allocating classical channels around the CV-QKD
channel so as to minimize SpRS; such a QAWA allows more
classical channels to be allocated without impacting CV-QKD traffic.

\subsubsection{Classical link congestion, effect of QKD-traffic margins on classical capacity}
\label{sec/results/margins}

In our model that allocates QKD first, after the full capacity
of a CV-QKD link is allocated, adding any optical power for
classical channels would push the SpRS noise above acceptable limits.
This link is then effectively ``clogged'' to classical traffic.
In the low-connectivity topology that we used, this can have an
adverse effect on network capacity, as even a small number of
unavailable links can isolate some nodes.
On the other hand, reserving a few-percent margin on QKD link capacity
during initial QKD traffic allocation allows all links to have at
least some SpRS tolerance, for a negligible impact on global QKD
capacity, but strongly reducing the impact of coexistence on classical
capacity.
Combining margins with QAWA, we observed scenarios where coexistence
had virtually no effect on classical traffic
(Fig.~\ref{fig/our-study}(c), \cite{garbhapu-ecoc-2024}).

\section{Conclusion and perspectives}
\label{sec/conclusion}

The integration of CV-QKD in classical networks presents both physical
and network-planning challenges. While coexistence is feasible, it requires
careful management of quantum communication devices, joint
optimization of classical and quantum traffic, and advanced planning
tools. Our simulations demonstrate that optimizing the placement of
CV-QKD channels and reserving margins on QKD link capacity can
significantly reduce the impact on classical network
capacity. However, further studies are needed to refine these
strategies, particularly in networks with more complex topologies
and especially varying node connectivity.
Also, the accuracy of the SpRS model will be critical to optimize
QKD channel placement.
Finally, these joint optimization strategies should be extended
beyond QKD to entanglement networks, using adapted routing
such as \cite{zeng-TNet-2024-entanglement-routing}, and measured
against theoretical capacity bounds such as
\cite{pirandola-Nature-2019-capacities-q-network,vardoyan-QCE-2023-q-network-utility,pouryousef-QuNet-2023-q-network-planning}.
As quantum technologies continue to advance,
addressing these network-level challenges will be critical to realizing
the full potential of quantum communication networks.

\section*{Acknowledgment}

This project has received funding from the European Union's Horizon
Europe research and innovation programme under the project ``Quantum
Security Networks Partnership'' (QSNP, grant agreement No 101114043).

\bibliography{IEEEabrv,ware-q-aware-network}
\bibliographystyle{utphys}

\end{document}